\documentclass[sigconf]{acmart}
\usepackage[normalem]{ulem}
\useunder{\uline}{\ul}{}
\usepackage{graphicx}
\usepackage{subcaption}
\usepackage{amsmath}
\usepackage{enumitem}
\usepackage{xspace}
\usepackage{balance}
\theoremstyle{definition}

\newcommand{\modelname}{\textsf{GLTA}\xspace}
\AtBeginDocument{%
  }

\copyrightyear{2025}
\acmYear{2025}
\setcopyright{acmlicensed}\acmConference[WWW Companion '25]{Companion Proceedings of the ACM Web Conference 2025}{April 28-May 2, 2025}{Sydney, NSW, Australia}
\acmBooktitle{Companion Proceedings of the ACM Web Conference 2025 (WWW Companion '25), April 28-May 2, 2025, Sydney, NSW, Australia}
\acmDOI{10.1145/3701716.3715583}
\acmISBN{979-8-4007-1331-6/25/04}

\begin{document}

\title{Training Large Recommendation Models via Graph-Language Token
Alignment}

\author{Mingdai~Yang}
\email{myang72@uic.edu}
\affiliation{%
  \institution{University of Illinois at Chicago}
  \city{Chicago}
  \country{USA}}

\author{Zhiwei~Liu}
\email{zhiweiliu@salesforce.com}
\affiliation{%
  \institution{Salesforce AI Research}
  \city{Palo Alto}
  \country{USA}
}

\author{Liangwei~Yang}
\email{liangwei.yang@salesforce.com}
\affiliation{%
  \institution{Salesforce AI Research}
  \city{Palo Alto}
  \country{USA}}

\author{Xiaolong~Liu}
\email{xliu262@uic.edu}
\author{Chen~Wang}
\email{cwang266@uic.edu}
\affiliation{%
  \institution{University of Illinois at Chicago}
  \city{Chicago}
  \country{USA}}

\author{Hao Peng}
\orcid{0000-0003-0458-5977}
\email{penghao@buaa.edu.cn}
\affiliation{%
   \institution{Beihang University,~\& Hangzhou Innovation Institute of BUAA}
   \country{Beijing~\& Hangzhou, China}}
\authornote{Corresponding author}

\author{Philip S.~Yu}
\email{psyu@uic.edu}
\affiliation{%
  \institution{University of Illinois at Chicago}
  \city{Chicago}
  \country{USA}}
\renewcommand{\shortauthors}{Mingdai Yang et al.}


\begin{abstract}
Recommender systems (RS) have become essential tools for helping users efficiently navigate the overwhelming amount of information on e-commerce and social platforms. However, traditional RS relying on Collaborative Filtering (CF) struggles to integrate the rich semantic information from textual data. Meanwhile, large language models (LLMs) have shown promising results in natural language processing, but directly using LLMs for recommendation introduces challenges, such as ambiguity in generating item predictions and inefficiencies in scalability. In this paper, we propose a novel framework to train Large Recommendation models via \textbf{G}raph-\textbf{L}anguage \textbf{T}oken \textbf{A}lignment. By aligning item and user nodes from the interaction graph with pretrained LLM tokens, \textbf{\modelname} effectively leverages the reasoning abilities of LLMs. Furthermore, we introduce Graph-Language Logits Matching (GLLM) to optimize token alignment for end-to-end item prediction, eliminating ambiguity in the free-form text as recommendation results.
Extensive experiments on three benchmark datasets demonstrate the effectiveness of \modelname, with ablation studies validating each component. 
\end{abstract}




\maketitle

\section{Introduction}
With the development of e-commerce and social platforms, information collection and decision-making have become essential yet overwhelming for individual customers. RS plays a key role in simplifying these tasks by offering personalized suggestions, and the recent successes of Graph Neural Networks (GNNs) have been developed to learn RS from graphs~\cite{LightGCN, hccf}.
Despite their effectiveness, graph-based RS often struggle to integrate rich semantic information from textual data, limiting their ability to capture nuanced user preferences and item characteristics. Leveraging large language models (LLMs) can bridge this gap by enhancing understanding of textual data and improving recommendation accuracy~\cite{rlmrec, ihp}.

A straightforward idea is to train an LLM on recommendation data to serve as a recommender~\cite{p5}.
While previous studies have demonstrated that LLMs can function as recommenders~\cite{cao-etal-2024-aligning}, they overlook that established graph-based recommendation models effectively leverage user-item interactions for collaborative filtering~\cite{UPRTH, GTGS}.
Moreover, when the LLM generates language tokens as item predictions, the output free-form text introduces ambiguity when matching the actual items, compared to traditional recommendation models that output a clear list of item IDs. 
To mitigate this, additional post-processing is needed to parse text back to specific items, which hinders the actual performance when candidate items are similar. 
Instead of deploying LLMs directly as RS, some recent works integrate LLM embeddings as supplementary features to enhance existing recommendation models~\cite{rlmrec,llmrec}. 
However, the recommendation process in their approaches is primarily driven by the graph-based model, leaving the reasoning capabilities of LLMs underutilized.

\begin{figure}[]
    \centering
    \includegraphics[scale=0.6]{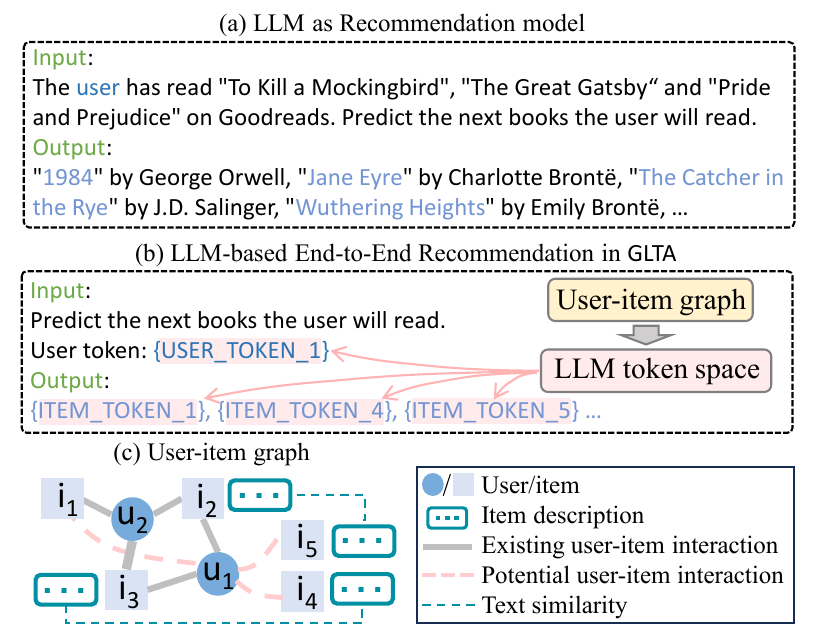}
    \
    \caption{A toy example of the input and the output in \modelname, compared with directly deploying LLM for recommendation.}
    \label{fig:overview}
\end{figure}

To effectively harness the powerful reasoning capability of LLMs for recommendation, 
we propose a novel framework, training large recommendation models via \textbf{G}raph-\textbf{L}anguage \textbf{T}oken \textbf{A}lignment, which aligns LLMs with graphs using a carefully designed token alignment paradigm.
To be concrete, we first align items with their text descriptions to obtain item tokens, and then align users with pretrained item and text tokens for recommendation. 
To implement end-to-end recommendation, we design a GLLM layer to optimize the token alignment by matching predicted item logits with ground-truth items. This GLLM layer eliminates the hallucination issue that non-existing items are generated when directly using LLM outputs as item prediction results. As shown in Figure~\ref{fig:overview}, different from directly adopting LLMs as recommendation models, our \modelname accurately generates existing item tokens instead of plain text outputs.
The key contributions of this paper are:
\begin{itemize}[leftmargin=*]
    \item We propose a novel recommendation framework, \modelname,  integrating LLMs and recommendation in an end-to-end manner, where each output of the LLM corresponds precisely to an item in RS, eliminating the hallucination issue and the ambiguity from free-form text as output. 
    \item We adaptively align nodes pretrained on the graph with pretrained tokens of the LLM by token projectors, and introduce a novel GLLM layer for optimizing end-to-end recommendation based on these aligned node tokens. Only projectors and GLLM layers are finetuned for this efficient end-to-end prediction.
    \item To verify the preeminence of \modelname, we conduct extensive experiments on three publicly available benchmark datasets. The effectiveness of each component in \modelname is verified through ablation studies.
\end{itemize}

\section{Preliminary}\label{sec:pre}
\subsection{Recommendation Task} Given two disjoint node sets, including a user set $\mathcal{U}$ and an item set $\mathcal{I}$, and the interactive edges, i.e., user-item edges $E_{\mathcal{U},\mathcal{I}}$, an interaction graph is defined as $\mathcal{G}=(V,E)$ where $V = \mathcal{U} \cup \mathcal{I}$. Besides, each item $i$ has a text description.
An end-to-end recommendation task for a user $u$ is to predict a ranking list of items $\{i_1, i_2, \dots, i_m\}$, with which this user has no interactions in the graph $\mathcal{G}$.
\subsection{Graph Pretraining}
Graph-based CF enhances recommendation by using graph structures to model user-item interactions.  In this work, we use LightGCN~\cite{LightGCN} as a graph-based CF method to capture and encode structure information on the user-item graph. In the first stage, user node embeddings and item node embeddings are pretrained as $\mathbf{E}_u\in\mathbb{R}^{|\mathcal{U}|\times d}$ and $\mathbf{E}_i\in\mathbb{R}^{|\mathcal{I}|\times d}$ and frozen in the following alignment stages, where $d$ denotes the dimension of pretrained node embeddings.
\section{Proposed Framework: \modelname}\label{sec:model}

\begin{figure*}[]
    \centering
    \includegraphics[scale=0.52]{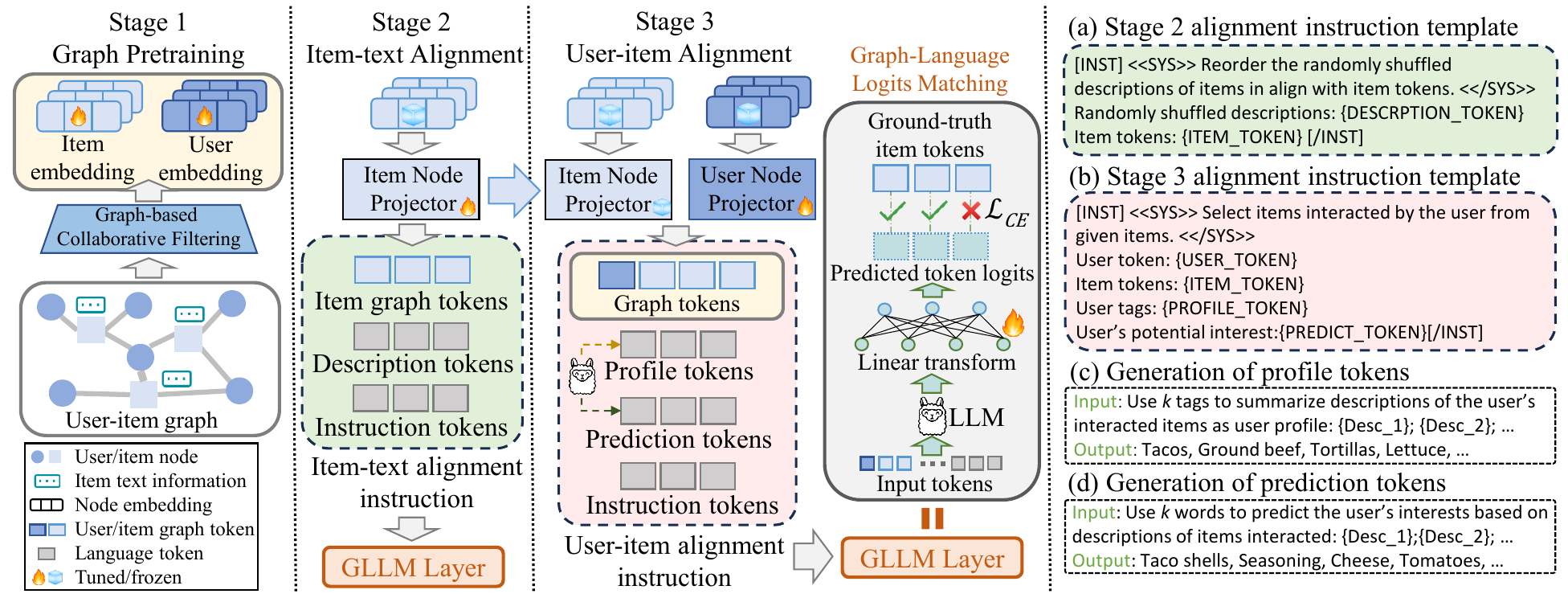}
    \caption{The framework of \modelname consists of three stages: Graph Pretraining, Item-text Alignment, and User-item Alignment. The instruction templates are shown on the right of the figure with the generation process of profile and prediction tokens.}
    \label{fig:framework}
\end{figure*}

\subsection{Item-text Alignment}
The proposed \textbf{\modelname} is shown in Figure~\ref{fig:framework}. Following graph pretraining, it is essential to align item node embeddings $\mathbf{E}_i$ with item descriptions tokens pretrained from the LLM. These two types of embeddings typically reside in different spaces. The item node embeddings capture collaborative signals from the graph, while the text embeddings capture semantic information from the textual descriptions. Inspired by GraphGPT~\cite{graphgpt}, we apply a simple linear layer as an item node projector that maps these item nodes into the same language token space with the descriptions of these items:
\begin{equation}\label{eq:itemalign}
    \mathbf{V}_i = \mathbf{W}_i\mathbf{E}_i+\mathbf{b}_i,
\end{equation}
where $\mathbf{V}_i\in\mathbb{R}^{|\mathcal{I}|\times d}$ is the embeddings of item tokens. $\mathbf{W}_i$ and $\mathbf{b}_i$ denote the weight and bias of the item node projector. Then, an item-text alignment instruction template, shown in Figure~\ref{fig:framework} (a), queries the LLM to reorder the item language information and match the token with the predicted logits.

\subsection{User-item Alignment}
Besides the item-text alignment, a user-item alignment is proposed to allow the LLM to process both user and item information in the same context. Similarly, a linear layer is used as a user node projector to map user nodes into the LLM token space.
\begin{equation}\label{eq:useralign}
    \mathbf{V}_u = \mathbf{W}_u\mathbf{E}_u+\mathbf{b}_u,
\end{equation}
where $\mathbf{V}_u\in\mathbb{R}^{|\mathcal{U}|\times d}$ is the embeddings of user tokens. $\mathbf{W}_u\in\mathbb{R}^{d\times d}$ and $\mathbf{b}_u\in\mathbb{R}^{1\times d}$ denote the weight and bias of the user node projector. 
This projector establishes the correspondence between the user nodes, the
language tokens and the item tokens pretrained in previous item-text alignment.

Besides these user tokens and pretrained item tokens, we introduce profile tokens and prediction tokens into the user-item alignment instruction template. These profile and prediction tokens are generated by the LLM based on item descriptions. The instruction templates for generating profile and prediction tokens are shown in Figure~\ref{fig:overview}(c) and Figure~\ref{fig:overview}(d), respectively. In this way, user node embeddings are mapped to the same token space with semantic characteristics reflecting their historical interactions and potential preferences. Then, the user-item alignment instruction template is fed into the LLM to generate the predicted item tokens for each user. During training, the item token prediction is optimized by GLLM for end-to-end recommendation. 

\subsection{Graph-Language Logits Matching}\label{sec:GLLM}
In a traditional setup when using LLMs as RS, additional steps are required to parse the text and map it back to specific items~\cite{graphgpt,p5}.
To eliminate the ambiguity that arises when interpreting free-form text outputs of the LLM, a GLLM layer is designed for the end-to-end recommendation in \modelname. After inputting the instruction template to the LLM, a linear layer is applied to transform the last-layer hidden states of the LLM into item token logits $\mathbf{Z}_{i}^u\in \mathbb{R}^{L\times |\mathcal{I}|}$, where $L$ denotes the maximum sequence length in the LLM. Then, a cross-entropy loss is applied to match the predicted logits to the ground-truth items interacted by the user:
\begin{equation}
    \mathcal{L}_{CE} = -\frac{1}{L} \sum_{t=1}^{L} \log \left( \frac{\exp\left( \mathbf{Z}_i[t, y_{i,t}^+] \right)}{\sum_{j=1}^{|\mathcal{I}|} \exp\left( \mathbf{Z}_i[t, j] \right)} \right),
\end{equation}
where $y_{i,t}^+$ is the ground-truth item ID at position $t$ in the sequence, and $\mathbf{Z}_i[t, y_{i,t}^+]$ represents the logit corresponding to the ground-truth item at position $t$. This cross-entropy loss provides direct supervision by comparing the model's predicted probability distribution against the actual ground-truth item IDs. In real-world datasets, the number of items interacted by the user is not necessarily equal to $L$, and the order information of the interacted item $y_{i,t}^+$ can be unavailable. In that case, we only optimize item token logits in the first $k$ positions, according to $k$ ground-truth items randomly shuffled from all the items interacted by the user. In this work, we use a quantized version of LLaMA-2-7B~\footnote{https://huggingface.co/TheBloke/Llama-2-7B-GPTQ} as the LLM for training efficiency and adopt Adam~\cite{Adam} as the optimizer.

\section{Experiment}
\subsection{Experiment Settings}
\subsubsection{Datasets}
\begin{table}
  \caption{Statistics of the Datasets}
  \centering
  \label{tab:datasets}\resizebox{0.45\textwidth}{!}{
  \begin{tabular}{l c c c}
        \hline
        \hline
        \textbf{Dataset} & Goodreads & Amazon & MovieLens \\
        \hline
        \textbf{\#Users} & 10,131 & 1,032 & 6,040\\

        \textbf{\#Items} & 10,725 & 1,7609 & 3,706\\

        \textbf{\#U-I interactions} & 478,334 & 30,510 & 1,000,209\\
        \textbf{Density} & 0.440\% & 0.168\% & 4.468\% \\
        \hline
  \end{tabular}}
\end{table}
We conduct experiments on three publicly available datasets: Goodreads~\cite{dataset:goodreads}, Amazon~\cite{dataset:amazon} and MovieLens\footnote{https://grouplens.org/datasets/movielens/1m/}.
We use history books and groceries as items with their corresponding descriptions in Goodreads and Amazon datasets, respectively. For MovieLens, we regard movies as items and movie genres as item descriptions since no movie descriptions are provided in this dataset. The details of the datasets are shown in Table~\ref{tab:datasets}.

\subsubsection{Baselines}
To demonstrate the effectiveness of \modelname, we compare it with three groups of representative baselines. 1.) GNN-based recommendation with only interaction information (LightGCN~\cite{LightGCN}, HCCF~\cite{hccf}) that applies graph or hypergraph neural network on the user-item interaction graph for information propagation. 2.) GNN-based recommendation with text information (LightGCN+, HCCF+) that uses InfoNCE loss to align the encoded node embeddings with description embeddings from a sentence transformer~\cite{minilm}. 3.) LLM-based recommendation~\cite{rlmrec} that leverages contrastive (RLMRec-Con) or generative (RLMRec-Gen) alignment. 

\subsubsection{Evaluation Metrics}
We evaluate the recommendation in an end-to-end manner by ranking the test users with all non-interacted items. Precision (P$@5$, P$@10$) and NDCG (N$@5$, N$@10$) are adopted as evaluation metrics.   

\begin{table*}[]\caption{Overall performance comparison. The best and second-best methods are in boldface and underlined.}\label{tab:overall}
\
\resizebox{0.95\textwidth}{!}{
\begin{tabular}{l|rrrr|rrrr|rrrr}
\hline
\hline
Dataset    & \multicolumn{4}{c|}{Goodreads}                                                                                         & \multicolumn{4}{c|}{Amazon}                                                                                            & \multicolumn{4}{c}{MovieLens}                                                                                           \\ \hline
Metric     & \multicolumn{1}{c}{P@5}     & \multicolumn{1}{c}{P@10}    & \multicolumn{1}{c}{N@5}     & \multicolumn{1}{c|}{N@10}    & \multicolumn{1}{c}{P@5}     & \multicolumn{1}{c}{P@10}    & \multicolumn{1}{c}{N@5}     & \multicolumn{1}{c|}{N@10}    & \multicolumn{1}{c}{P@5}     & \multicolumn{1}{c}{P@10}    & \multicolumn{1}{c}{N@5}      & \multicolumn{1}{c}{N@10}     \\ \hline
LightGCN   & 0.1999                      & 0.1618                      & 0.2346                      & 0.2352                       & 0.0204                      & 0.0139                      & 0.0301                      & 0.0305                       & 0.0470                      & 0.0426                      & 0.0463                       & 0.0466                       \\
HCCF       & 0.1998                      & 0.1632                      & 0.2371                      & 0.2439                       & 0.0209                      & 0.0143                      & 0.0275                      & 0.0304                       & 0.0467                      & 0.0419                      & 0.0482                       & 0.0486                       \\
LightGCN+  & 0.2004                      & 0.1616                      & 0.2347                      & 0.2358                       & 0.0211                      & 0.0150                      & 0.0268                      & 0.0290                       & {\ul 0.0527}                & 0.0451                      & 0.0467                       & 0.0469                       \\
HCCF+      & 0.2034                      & 0.1641                      & 0.2379                      & 0.2483                       & 0.0217                      & {\ul 0.0156}                & 0.0310                      & 0.0315                       & 0.0516                      & 0.0436                      & 0.0486                       & 0.0488                       \\
RLMRec-Con & 0.2062                      & {\ul 0.1676}                & {\ul 0.2441}                & {\ul 0.2542}                 & {\ul 0.0227}                & 0.0142                      & {\ul 0.0348}                & {\ul 0.0358}                 & 0.0499                      & {\ul 0.0460}                & {\ul 0.0488}                 & {\ul 0.0492}                 \\
RLMRec-Gen & {\ul 0.2062}                & 0.1653                      & 0.2407                      & 0.2424                       & 0.0223                      & 0.0145                      & 0.0336                      & 0.0339                       & 0.0517                      & 0.0451                      & 0.0475                       & 0.0477                       \\ \hline
\modelname      & \textbf{0.2465}             & \textbf{0.1882}             & \textbf{0.2722}             & \textbf{0.2905}              & \textbf{0.0359}             & \textbf{0.0236}             & \textbf{0.0499}             & \textbf{0.0500}              & \textbf{0.0968}             & \textbf{0.0821}             & \textbf{0.1162}              & \textbf{0.1403}              \\ \hline
Improv.    & \multicolumn{1}{l}{19.54\%} & \multicolumn{1}{l}{12.29\%} & \multicolumn{1}{l}{11.51\%} & \multicolumn{1}{l|}{14.28\%} & \multicolumn{1}{l}{58.14\%} & \multicolumn{1}{l}{51.28\%} & \multicolumn{1}{l}{43.39\%} & \multicolumn{1}{l|}{39.66\%} & \multicolumn{1}{l}{83.68\%} & \multicolumn{1}{l}{78.48\%} & \multicolumn{1}{l}{138.11\%} & \multicolumn{1}{l}{185.16\%} \\ \hline
\end{tabular}
}
\end{table*}

\subsection{Overall Performance}
Performance comparison between \modelname and other baselines are shown in Table~\ref{tab:overall}. We have the following observations. First,  \modelname exhibits superior performance on all datasets, especially in MovieLens where the dense user-item interactions lead to over-smoothing in GNN-based baselines~\cite{oversmooth}. Second, compared to graph-based recommendation with only interaction information, introducing text information into graph-based methods improves the recommendation performance in most cases, which verifies the importance of incorporating rich semantic content to enhance user-item matching and better capture the contextual nuances of items. Third, the LLM-based recommendation method, RLMRec-Con, has better overall performance than other baselines. This justifies that aligning the knowledge of LLMs with collaborative
relation learning through contrastive learning is able to enhance recommendation performance. However, the recommendation process is still done by the backbone LightGCN model in RLMRec, leaving the reasoning abilities of LLMs untouched. On the contrary, \modelname projects users and items into language space as tokens first, and then completely leverages the LLM for end-to-end recommendation, which is a more holistic approach that fully utilizes the reasoning capabilities and contextual understanding of LLMs.

\subsection{Item-text and User-item Alignment}
\begin{figure}[]
    \begin{subfigure}{0.153\textwidth}
    \includegraphics[width=\textwidth]{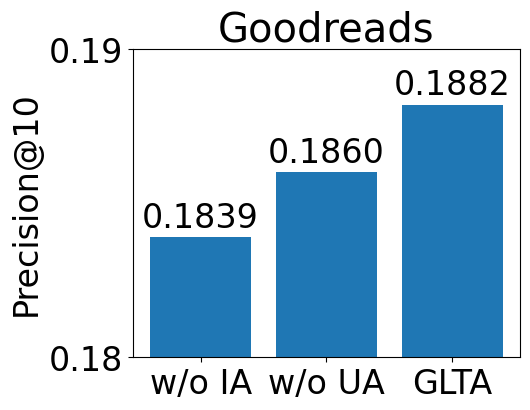}
    \end{subfigure}
    \hfill
    \begin{subfigure}{0.153\textwidth}
    \includegraphics[width=\textwidth]{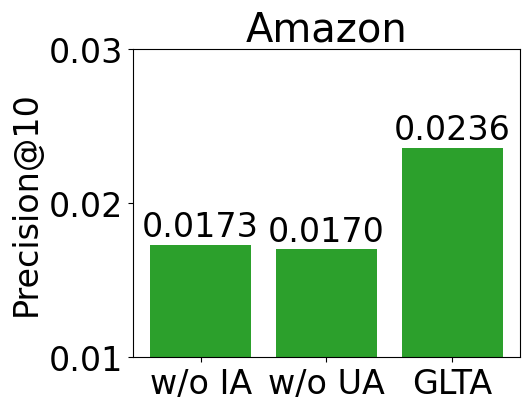}
    \end{subfigure}
    \hfill
    \begin{subfigure}{0.153\textwidth}
    \includegraphics[width=\textwidth]{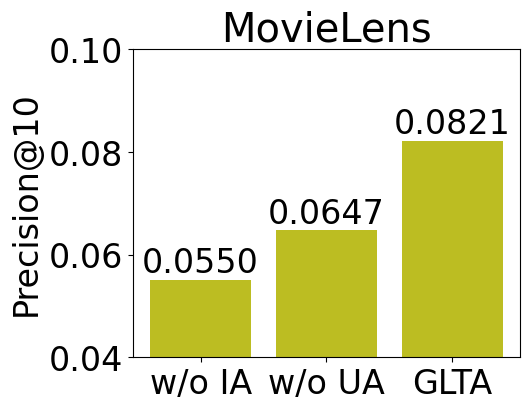}
    \end{subfigure}\
    \
        \caption{Performance of \modelname compared to its variants without item-text alignment and without user-item alignment.}\label{fig:ablation_1}
\end{figure}

In \modelname, item-text alignment and user-item alignment are designed to align CF with the reasoning ability of the LLM. To verify the effectiveness of these two alignment methods, we compare the performance of \modelname to its variant without item-text alignment (w/o IA) and without user-item alignment (w/o UA) on the three datasets in Figure~\ref{fig:ablation_1}. For the variant \textit{w/o IA}, we directly remove the item-text alignment stage and update item and user node projectors simultaneously in the user-item alignment stage. For the variant \textit{w/o UA}, we use user IDs instead of aligned user tokens in the user-item alignment instruction template before feeding it into the GLLM layer. We find that employing item-text alignment or user-item alignment leads to stable enhancement in all datasets, which implies that both alignment methods are crucial for effectively leveraging the LLM's reasoning capabilities in the recommendation process.

\subsection{Profile and Prediction Tokens}
To quantify the contribution of profile and prediction tokens used in user-item alignment, we conduct an ablation study to investigate the performance of \modelname without these tokens generated by the LLM. The results are shown in Table~\ref{tab:ablation_2}. Both profile and prediction tokens generally improve performance if included in the instruction template. Notably, the advantages of using these LLM-generated tokens become more pronounced when the distinctions between items in the dataset are clearer. For instance, the performance improvement is more evident in MovieLens, where items span a wide range of movie genres, but less pronounced in Goodreads, where items consist only of history books. We speculate that the broader diversity of items in datasets like MovieLens allows the LLM-generated tokens to better capture user features from nuanced differences between items, thereby enhancing recommendation accuracy. In contrast, as the distinctions between items are less varied and therefore less reliant on the LLM's enhanced representation capabilities, the additional benefits provided by these tokens are also limited.
\begin{table}[]\caption{Performance of \modelname compared to its three variants without profile (PF) tokens and/or without prediction (PD) tokens on three datasets.}\label{tab:ablation_2}
\
\resizebox{0.48\textwidth}{!}{
\begin{tabular}{l|cc|cc|cc}
\hline
\multicolumn{1}{c|}{Dataset} & \multicolumn{2}{c|}{Goodreads}    & \multicolumn{2}{c|}{Amazon}       & \multicolumn{2}{c}{MovieLens}     \\ \hline
Metric                       & P@5             & N@5             & P@5             & N@5             & P@5             & N@5             \\ \hline
w/o both                     & 0.2462          & 0.2705          & 0.0316          & 0.0408          & 0.0736          & 0.0658          \\
w/o PF                       & 0.2466          & 0.2668          & 0.0318          & 0.0414          & 0.0763          & 0.0713          \\
w/o PD                       & \textbf{0.2467}          & 0.2688          & 0.0341          & 0.0449          & 0.0860          & 0.0761          \\ \hline
\modelname                        & 0.2465 & \textbf{0.2722} & \textbf{0.0359} & \textbf{0.0499} & \textbf{0.0968} & \textbf{0.1162} \\ \hline
\end{tabular}
}
\
\end{table}

\subsection{Item Prediction in GLLM}
\begin{figure}[]
    \begin{subfigure}{0.153\textwidth}
    \includegraphics[width=\textwidth]{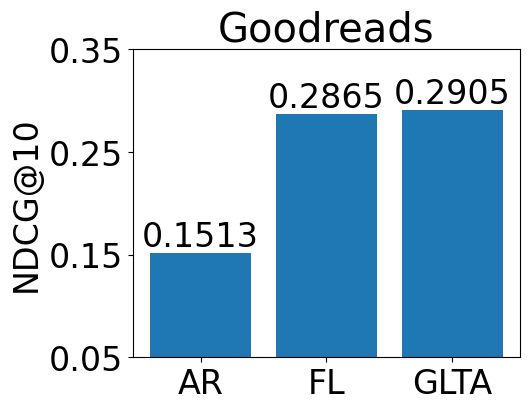}
    \end{subfigure}
    \hfill
    \begin{subfigure}{0.153\textwidth}
    \includegraphics[width=\textwidth]{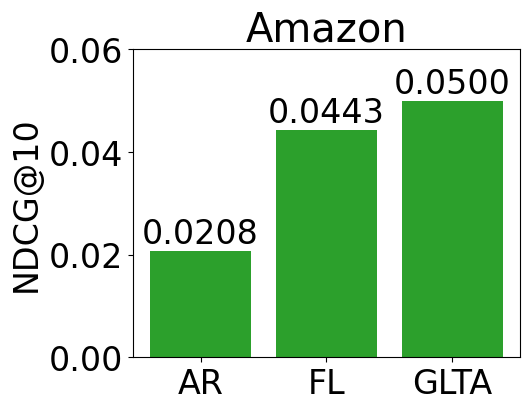}
    \end{subfigure}
    \hfill
    \begin{subfigure}{0.153\textwidth}
    \includegraphics[width=\textwidth]{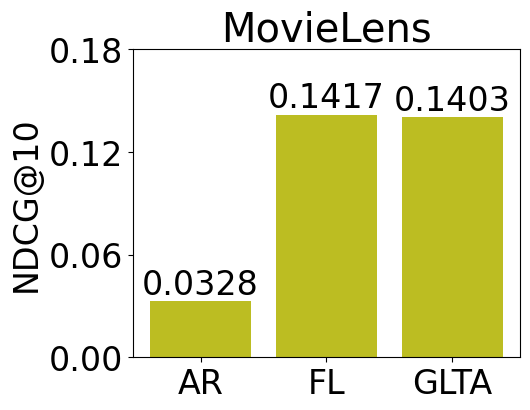}
    \end{subfigure}
        \caption{Performance of \modelname compared to its variants with different optimization methods on three datasets.}\label{fig:ablation_3}
\end{figure}

We further explore two other item prediction patterns in GLLM layers. Autoregressive inference(AR): During inference, we strictly follow the text generation of LLMs in which each item token is generated based on the previous item tokens the model has generated. First-logit optimization (FL): To predict the top $k$ favorite items for each user, we use the largest $k$ elements in the first logit from $\mathbf{Z}_{i}^u\in \mathbb{R}^{L\times |\mathcal{I}|}$, instead of the first $k$ logits in \modelname. The results are demonstrated in Figure~\ref{fig:ablation_3}. The poor performance of autoregressive inference indicates that, unlike natural language processing tasks where contextual continuity is vital, the sequential dependency between items is not as strong in recommendation. The inconsistency between training and inference patterns exacerbates this issue, particularly because the model is trained using item descriptions and collaborative signals rather than direct sequential user-item interaction data.

\section{Conclusion}\label{sec:conclusion}
In this paper, we propose a novel recommendation framework \modelname, which applies token alignment to integrate graph-based CF with the reasoning capabilities of LLMs. 
In \modelname, we begin by employing a graph encoder to capture user and item node features from the collaborative relationships within the user-item graph. Next, item and user node embeddings are adaptively aligned with other language tokens using node projectors. Finally, the user and item node projectors are optimized for end-to-end recommendation through the GLLM layer. 
Compared with finetuning LLMs as RS, \modelname overcomes hallucination and is efficient since only projectors and GLLM layers are finetuned. Future works may explore integrating additional modalities with graph-based CF through multimodal large language models.

\section{Acknowledgments}
This work is supported in part by NSF under grants III-2106758, and POSE-2346158.
Hao Peng is supported by the National Key R\&D Program of China through grant 2022YFB3104703, the NSFC through grants 62322202 and 62441612, Local Science and Technology Development Fund of Hebei Province Guided by the Central Government of China through grant 246Z0102G, the "Pioneer” and “Leading Goose” R\&D Program of Zhejiang" through grant 2025C02044, Hebei Natural Science Foundation through grant F2024210008, and the Guangdong Basic and Applied Basic Research Foundation through grant 2023B1515120020.
\bibliographystyle{ACM-Reference-Format}
\balance
\bibliography{LRTA}

\end{document}